\documentclass[12pt,twoside]{article}
\usepackage{a4wide}

\usepackage{amssymb}
\usepackage{amsmath}

\usepackage{setspace}
\author{V.~A.~Malyshev 
\and A.~D.~Manita\thanks{Moscow State University, Faculty of Mechanics
and Mathematics, Leninskie Gory 1, GSP-1,
 Moscow, 119991, Russia; e-mail: malyshev2@yahoo.com, e-mail: manita@mech.math.msu.su}}

\date{ }

\title{Stochastic micromodel of the Couette flow}

\begin{document}

\newtheorem{theorem}{Theorem}
  \newtheorem{lemma}[theorem]{Lemma}
  \newtheorem{corollary}[theorem]{Corollary}
  \newtheorem{proposition}[theorem]{Proposition}
  \newtheorem{definition}[theorem]{Definition}

\newcommand{\ds}{\displaystyle}
\newcommand{\vstir}[2]{v^{(#1)\leftrightarrow (#2)}}
\newcommand{\indik}{{\mathbf 1}}
\newcommand{\ncc}{\varnothing }
\newcommand{\er}{b}
\newcommand{\auxpr}{\vartheta }
\newcommand{\zet}{\zeta }
\newcommand{\const}{\,\mbox{const}\,}
\newcommand{\No}{N.}

\newcommand{\sh}{\sinh}

\newcommand{\otk}{``\ignorespaces}
\newcommand{\zaks}{''\ }
\newcommand{\zak}{''}

\newcommand{\bZ}{\mathbf{Z}}
\newcommand{\bR}{\mathbf{R}}
\newcommand{\ccW}{\mathcal{W}}
\newcommand{\ccL}{\mathcal{L}}
\newcommand{\bP}{\mathbf{P}}

\newcommand{\kb}{{\upshape, }}
\newcommand{\kbt}{{\upshape; }}
\newcommand{\dvt}{{\upshape: }}

\newcommand{\ccM}{\mathcal{M}}
\newcommand{\Reop}{\mathop{\rm Re\,}}
\newtheorem{lemm}{Lemma}
\newcommand{\Proof}{{\em Proof}. }
\newcommand{\cl}{\colon\,}

\newcommand{\remno}[1]{{\em Remark\/ }#1. }

\catcode`\@=11
\def\SimpleEqNumbers{\def\cl@section{}\@addtoreset{subsection}{section}
\renewcommand{\theequation}{\arabic{equation}}}
\makeatother

\maketitle    

\markboth{\null\hfill\hfill Stochastic Micromodel Of The Couette Flow}{{Malyshev and Manita\hfill\hfill\null}}

\begin{abstract}
We study Markov exclusion process for a particle system with 
a local interaction in the integer  strip. This process models the 
exchange of velocities and particle-hole exchange of the liquid 
molecules. 
It is shown that the mean velocity profile  corresponds to 
the behaviour which is characteristic for incompressible   viscous 
liquid. We prove the existence of phase transition between laminar
and turbulent profiles.
\end{abstract}

{\small 
\par}

\pagestyle{myheadings} 
\thispagestyle{plain}  

\setcounter{page}{1}

\section{Introduction} 
From the mathematical point of view this paper considers a phase 
transition for a Markov multi-dimensional  exclusion process. But 
as this problem originated from concrete physical models, we discuss
these connections in detail. 

One can study the liquid flow from macroscopic and from microscopic
point of view. Although in the first case (that is in the continuum 
mechanics) most studies are  based on the Navier-Stokes equations, there 
exist various generalizations introducing randomness in 
these equations (see the review of J. Marsden and other papers in
~\cite{Strange}, some history of the question one can find in 
the recent preprint ~\cite{Shir}). It is precisely with 
Navier-Stokes equations the immense number of concrete theoretical 
and practical problems of the liquid mechanics, including the Couette
flow, which is the simplest (but far from being trivial model, see 
~\cite{Kolmogor},~\cite{MeshSin}) available for a mathematical
study. 

However, there are only very few microscopic models for concrete 
situations with nonzero viscosity, in despite of the abundance of
lattice models with infinite number of particles 
(see books ~\cite{Spohn},~\cite{MarPul},~\cite{Kipnis}). In theoretical physics the 
preferable direction is the derivation of the Navier-Stokes equations 
following the sequence: hamiltonian microdynamics  $\to$ BBGKY 
equations  $\to$ Boltzmann equation   $\to$ equation of  
hydrodynamics. However, there is also direct derivation of the
Navier-Stokes equations from stochastic dynamics of particles 
on the lattice (see ~\cite{EsMaYao}). 

Recently, there appeared several papers where some particle system 
were related to the Navier-Stokes equations so that for a certain 
scaling the dynamics of these particle system converges to these 
equations (see ~\cite{MarPul},~\cite{Meleard},~\cite{Fontbona}).
However, in these papers the particle dynamics is not local, it is
close to the mean-field dynamics of McKean-Vlasov type. But most
important is that it is derived not from proper molecular dynamics but
from the computer simulation of Navier-Stokes equations, that is from
the Navier-Stokes equations themselves. 

The conclusion is that the 
derivation of Navier-Stokes equations from physically grounded molecular 
dynamics is still open. The main problem here is that it is far
from clear how to describe the molecular dynamics. The model of hard
balls, obeying the laws of classical mechanics, is simple in
formulation but very difficult to study. Moreover, one cannot be
sure that this model is adequate. The question is that the 
molecules in a liquid are very close to each other, and quantum
effects seem to be important. If one will work in the framework of
classical models, then one should take into account the interaction
of rotating and oscillating degrees of freedom of the neighbor 
molecules, which presumably have a random distributions.
 
From the other side, before any attempt to  derive  the 
Navier-Stokes equations, it could be natural first to try
particular cases. In the local model of the Couette flow considered 
below we use two physical processes: exchange of velocities between
molecules and the exchange between molecules and holes (that is 
empty places). This corresponds to the intuitive picture often 
discussed in the physical literature. Moreover, one can hope that 
there is a kind of universality, that is independence of qualitative
behaviour of the concretization of the model.

From probabilistic point of view we study a Markov system of locally
interacting particles, the dynamics of which is a mixture of two
well-known  (see ~\cite{Liggett}) exclusion processes - symmetric and completely asymmetric. 
Introducing boundary conditions does not 
influence the moment closeness of the process ~\cite{IMM}, and this
simplifies essentially the study. Precisely using moment closeness
one can avoid such difficult techniques as Bethe   
~\cite{Golinelli} or matrix  ~\cite{Derrida07} ansatz.

It is worth notice that, due to different scaling, direct study of 
concrete micromodel is not equivalent to the derivation of the 
Navier-Stokes equations and their subsequent application to the 
same situation.  

For our problem we get the phase transition for the velocity  
profile, corresponding to the transition in the Couette flow ~\cite{LanAxiLif}, 
however there is a difference.  The transition  parameter 
in our paper is not the classical Reynolds number, which depends 
on the viscosity, cross-section and longitudinal  velocity, but its analog, 
depending on the cross-exchange parameter (analog of 
viscosity), cross-section and random perturbation. The latter
could be the consequence of fast vortex, resulting from deviations
of velocity from longitudinal direction or from interaction of 
velocity with randomly distributed   internal degrees of freedom.
It appears interesting that the phase transition is sharp on the
rougher scale and smooth on the finer scale.

\section{Model and results}

The Couette flow is the liquid flow in one (horizontal) direction
in the space (which is discrete and two-dimensional, however
the generalization to multi-dimensional case is straightforward)
between two plates. One of the plates is fixed and the other
moves with constant velocity, that draws the neighbor particles 
with it. Other particles are also involved due to viscosity, that
is reflected microscopically in the exchange of velocities
between neighbor particles.

We come now to exact definitions. The basic set for us is the 
discrete strip 
$$
L_{S}=\{0,1,2,\ldots,S,S+1\}\times\bZ.
$$
To each point $(s,x)\in L_{S}$ we assign random variable $v_{(s,x)}$, 
which can take one of the following three values: $\ncc$, $0$ or $V$. 
The value
$\ncc$ means that there is no particle at the site  $(s,x)$  (\otk a hole\zak),
$v_{(s,x)}=0$ means 
the presence of one  particle with zero velocity,
and $v_{(s,x)}=V$ means the presence of one particle  with  velocity
 $V$. For notational convenience 
 the  \otk hole\zaks  will be called the particle with velocity $\ncc$.

{\it Dynamics} of this particle system consists of jumps and is 
defined by the following transitions. 
On any small time interval $[t,t+dt]$
 the following events can take place independently of one another:

{\rm (a)}~particles of each neighboring vertical pair  (at points  $(s,x)$ and $(s+1,x)$) 
exchange their velocities with probability $\lambda\, dt+o\,(dt)$ 
({\it velocity exchange between vertical layers}); 

{\rm (b)}~particles in each neighboring horizontal pair of the  type 
 $v_{(s,x)}=V$, $v_{(s,x+1)}=\ncc$ 
(and only of this type) exchange 
their velocities with probability $\lambda_{1}dt+o\,(dt)$ 
({\it horizontal flow}). Of course, one could put 
$V=\lambda_{1}$, but in this paper it will be convenient  to distinguish $V$ and $\lambda_{1}$.

{\rm (c)}~ on the zero layer each particle that has velocity $V$ acquires velocity  $0$ 
with probability $\beta\,dt+o\,(dt)$; on the top layer $S+1$ each particle 
that has velocity $0$ acquires velocity  $V$ 
with probability  $\beta\, dt+o\,(dt)$
({\it influence of the boundaries}); 

{\rm (d)}~each node  $(s,x)$ with nonempty velocity ($v_{(s,x)}=0$ or $V$)
 changes its velocity
 according to the rule $0\leftrightarrow V$  with probability  $\varepsilon\, dt+o\,(dt)$ 
 ({\it random perturbation}). 

In the sequel we assume that
 $\lambda>0$, $\lambda_1>0$, $\beta>0$ and $\varepsilon\ge 0$.

Thus we have just defined a continuous time {\it Markov} process $v(t)=\{ v_{(s,x)}(t),\ 
(s,x)\in L_{S}\}$, $t\in\bR_{+}$, with state space $\ccW=\{\ncc,0,V\}^{L_{S}}$. 
Its generator acts on the functions~$f$ 
that depend on velocities at finite number of nodes as follows
\begin{eqnarray}
\ccL f(v)&:=& \lambda\sum_{x\in\bZ}\sum_{k=0}^{S}\big(f(\vstir{k,x}{k+1,x})-f(v)\big)\nonumber\\
&& +\,\lambda_{1}\sum_{x\in\bZ}\sum_{k=0}^{S+1}\big(f(\vstir{k,x}{k,x+1})-f(v)\big)\,
\indik({v_{(k,x)}=V,v_{(k,x+1)}=\ncc})\label{eq:gen-L-obsch}\nonumber\\
&& +\,\beta\sum_{x\in\bZ}\big(f(v^{(0,x)})-f(v)\big)\,\indik({v_{(0,x)}=V})\nonumber\\
&&+\,\beta\sum_{x\in\bZ}\big(f(v^{(S+1,x)})-f(v)\big)\,\indik({v_{(S+1,x)}=0})\nonumber\\
&& +\,\varepsilon\sum_{x\in\bZ}\sum_{k=0}^{S+1}\big(f(v^{(k,x)})-f(v)\big)\,
\indik({v_{(k,x)}\not=\ncc}),
\end{eqnarray}
where two configurations $w=v^{y\leftrightarrow z}$ and $v$ from $\ccW$ differ 
only by the velocities at the points
 $y$ and $z$ of $L_{S}$:
$$
w_{y}=v_{z},\quad w_{z}=v_{y},
$$
and $v^{y}\in\ccW$  is a configuration that can differ from $v$ only in velocity value 
at node~$y$ due to the exchange $0\leftrightarrow V$:
$$
(v^{y})_{z}=\left\{\begin{array}{rl}
v_{z}, & \quad z\not=y,\\
\ncc, & \quad z=y,\quad v_{y}=\ncc,\\
V\,\indik({v_{y}=0}), & \quad z=y,\quad v_{y}\not=\ncc.\end{array}\right.
$$
One can check that such definition of $\ccL$ leads to a correct definition of the Markov process
 on $\ccW$ (see~\cite[Ch. I, \S 3]{Liggett}).

Denote by  $\eta^{t}$  the distribution of the stochastic process  $v(t)$ at time $t$.

Let  $A=((s_{1},x_{1}),\ldots,(s_{m},x_{m}))$ be some ordered finite subset of  $L_{S}$
and  $E=(e_{1},\ldots,e_{m})$ be 
some ordered collection of $e_{j}\in\{ \ncc,0,V\} $.
Consider the following probabilities
$$
q_{t}[A;E]:=\bP\,\{ v_{(s_{1},x_{1})}(t)=e_{1},\ldots,v_{(s_{m},x_{m})}=e_{m}\}.
$$

On the set $L_{S}$ we define the  action of the group  $(T_{a},\ a\in\bZ)$ of horizontal 
translations:
$$
T_{a}(s,x)=(s,x+a).
$$

In the present paper we restrict ourself to consideration of initial 
distributions $\eta^{0}$  of the process  $v$ that are {\it invariant}
 with respect to the action of the group~$T_{a}$.

Since the Markov semigroup generated by~(\ref{eq:gen-L-obsch}) commutes with the translations $T_{a}$,   for any time ~$t$ the distribution~$\eta^{t}$ is translation invariant. 
In particular, for any choice of sets~$A$ and $E$
\begin{equation}
q_{t}[T_{a}A;E]=q_{t}[A;E]\label{eq:m1-m2-tr-inv}
\end{equation}
for all $t>0$ and $a\in\bZ$. 
From physical point of view it means that we are interested only in homogeneous flows
that are invariant with respect to shifts along the  $x$-direction: $x\rightarrow x+a$.

As it is easy to see, the dynamics of the process $v$ is such that the total number of particles, 
namely, the  number of nodes with 
velocities  $0$ and $V$, is a conserved quantity.  
In particular, the mean (expected) number of nonempty nodes in a vertical section~$x$ 
is conserved in time and does not depend on~$x$:
\begin{equation}
\sum_{k=0}^{S+1}(q_{t}[(k,x);0]+q_{t}[(k,x);V])\equiv M\quad\forall x,t\,.\label{eq:M-eta-0}
\end{equation}
Evidently, the constant $M=M(\eta^{0})$ can be calculated in terms of the initial 
distribution~$\eta^{0}$. However, the mean number of nodes with velocity 
$V$ is not conserved.

From now on we are interested only in distributions in a fixed vertical layer,
therefore, via the homogeneity, the dependence on  $x$ will be omitted 
in the following notation
\begin{equation}
q_{t}[(s,x);e]=p_{s}^{e}(t).\label{eq:p-e-s-t}
\end{equation}

\begin{theorem}\label{t-first} Assume that the initial distribution
$\eta^{0}$ is translation invariant. Then\dvt

{\rm 1)}~$($existence of a stationary regime$)$ the following limits exist for any $s$ and $e$ 
$$
\mu_{s}^{e}=\lim_{t\rightarrow\infty}p_{s}^{e}(t),
$$
we call $\mu_{s}^{e}$ stationary probabilities\kbt

{\rm 2)}~$($uniformity of a particle density$)$ the 
stationary probabilities of {\rm\otk}holes{\rm\zak} are the same for
any layer $s=0,1,\ldots,S+1$\dvt
$$
\mu_{s}^{\ncc}\equiv\const.
$$
\end{theorem}

First we consider the case when the random perturbation is absent, i.e.,
  $\varepsilon=0$.

\begin{theorem}[{\rm linear velocity profile}]\label{t-eps-0}
Assume that the initial distribution   $\eta^{0}$ of the process is translation invariant and  $\varepsilon=0$.
Then the mean velocities of particles form a linear profile\kb  namely\kb 
$$
\mu_{k}^{V}=\frac{\rho_{S}}{S+1+2\lambda\beta^{-1}}\,(k+\lambda\beta^{-1}),\qquad k=0,1,\ldots,S+1,
$$
where $\rho_{S}=M(\eta^{0})/(S+2)$.
\end{theorem}

The velocity profile of this theorem corresponds to the laminar flow 
in contrast to what is observed in the turbulent regime
 (see~\cite{LanAxiLif}). In the next theorem we consider nonzero random perturbations.

\begin{theorem}\label{t-turb}
Assume that $S\rightarrow\infty,$ the parameter $\varepsilon=\varepsilon_{S}$
depends on $S$ and  the sequence of initial distributions
 $\{ \eta^{0,S}\}$ is such that the density of particles is fixed, i.e., 
\begin{equation}
\frac{M(\eta^{0,S})}{S+2}\rightarrow\rho\in[0,1],\qquad S\rightarrow\infty.\label{eq:plotnost}
\end{equation}
Then\dvt

{\rm 1)}~$($rough scale picture of the transition$)$ 
in scaling  $s=uS$ the functions  $\mu_{s}^{0}$ and $\mu_{s}^{V}$ demonstrate 
a phase transition from laminar profile to turbulent one\kbt  namely\kb 
if
 $\varepsilon_{S}S^{2}\rightarrow0,$
then the limiting profile of the function $\mu_{uS}^{V},$ $u\in(0,1),$ has a linear  $($laminar$)$
form\dvt 
$$
\mu_{uS}^{V}\rightarrow\rho u\qquad(S\rightarrow\infty),
$$
but, if  $\varepsilon_{S}S^{2}\rightarrow\infty,$ then the limiting profile is constant $($turbulent$)$\dvt
$$
\mu_{uS}^{V}\,\rightarrow\,\frac{\rho}{2},\qquad(S\rightarrow\infty);
$$
 
{\rm 2)}~$($finer scale picture of the transition$)$ 
the case $\varepsilon_{S}S^{2}=\const,$ $S\rightarrow\infty,$ is intermediate 
between the situations described in point~$1)$. Precisely, 
if we introduce a positive parameter  $K$   putting
$\varepsilon_{S}S^{2}=\frac{1}{2}\lambda K^{2},$ 
then
 $\mu_{uS}^{V}\rightarrow g_{K}(u),$ where
$$
g_{K}(u):=\frac{\rho}{2}\Big(1+\frac{\sh(K(u-1/2))}{\sh(K/2)}\Big).
$$
Since $\lim_{K\rightarrow0}g_{K}(u)=\rho u$ and
 $\lim_{K\rightarrow\infty}g_{K}(u)={\rho}/{2},$
we can say that by changing the parameter  $K>0$ we observe a smooth transition 
between laminar and turbulent profiles.
\end{theorem}

\remno{1} The parameter $K$ can be considered as  analog of the Reynolds number  $\Reop$. 
This can be easily explained by comparing these two numbers, that is by comparing the definition 
of  $K$ 
$$
K^{2}=\frac{S^{2}\varepsilon}{\lambda/2}
$$
with the definition of the Reynolds number
$$
\Reop=\frac{LV}{\nu},
$$
where $L$ is the cross-section of the flow, $V$ is the mean velocity, $\nu$ is the viscosity.

The mentioned phase transition is based on the competition between
processes which reinforce and the process which diminish the 
influence of the boundary. Thus, the factors $L$ and $S$ show that 
the greater the distance from the boundaries, the more  random 
perturbation diminishes the influence of the boundaries. The factors
$V$ and $\varepsilon$ show that if the velocity increases  the 
particles are perturbed more often, and the influence of the 
boundaries is less visible. Finally, when  $\lambda$ increases, then the influence of the boundaries increases. One should note that, by definition of the model,  the parameter $V$ can be put proportional to 
$\lambda_{1}$.

\section{Proofs}

\subsection{Closed equations for one-particle functions}
Our next task is to obtain a system of differential equations for the functions
 $p_{k}^{e}(t)$, $0\le k\le S+1$, $e\in\{\ncc,0,V\}$, defined in~(\ref{eq:p-e-s-t}).
First note that for any
 $k=0,1,\ldots,S+1$
\begin{equation}
p_{k}^{\ncc}(t)+p_{k}^{0}(t)+p_{k}^{V}(t)=1\quad\forall t\ge0,\label{eq:normQ0V}
\end{equation}
hence, for example, the functions $p_{k}^{0}(t)$ can be easily expressed 
in terms of the functions $p_{k}^{\ncc}(t)$ and $p_{k}^{V}(t)$.

\begin{lemm} \label{l-zamk-odnoch} 
Let the initial distribution
$\eta^{0}$ be translation invariant. 
Then the functions $p_{k}^{e}(t)$ satisfy to a closed system of linear 
first-order differential equations having the following structure.

{\rm 1)}~The functions $p_{k}^{\ncc}(t)$ can be  obtained from a subsystem
\begin{eqnarray}
\frac{d}{dt}p_{k}^{\ncc}(t) & = & \lambda(p_{k+1}^{\ncc}(t)+p_{k-1}^{\ncc}(t)-2p_{k}^{\ncc}(t)),\qquad 1\le k\le S,\nonumber \\
\frac{d}{dt}p_{0}^{\ncc}(t) & = & \lambda(p_{1}^{\ncc}(t)-p_{0}^{\ncc}(t)),\label{eq:evol-dyrok}\\
\frac{d}{dt}p_{S+1}^{\ncc}(t) & = & \lambda(p_{S}^{\ncc}(t)-p_{S+1}^{\ncc}(t)).\nonumber
\end{eqnarray}

{\rm 2)} The equations for $p_{k}^{V}(t)$ have the following form
\begin{eqnarray}
\frac{d}{dt}p_{k}^{V}(t) & = & \lambda(p_{k+1}^{V}(t)+p_{k-1}^{V}(t)-2p_{k}^{V}(t))+
\varepsilon(1-p_{k}^{\ncc}(t)-2p_{k}^{V}(t)),\nonumber\\ 
&&1\le k\le S,\nonumber\\
\frac{d}{dt}p_{0}^{V}(t) & = & \lambda(p_{1}^{V}(t)-p_{0}^{V}(t))+\varepsilon(1-p_{0}^{\ncc}(t)-2p_{0}^{V}(t))-\beta p_{0}^{V}(t),\label{eq:evol-V}\\
\frac{d}{dt}p_{S+1}^{V}(t) & = & \lambda(p_{S}^{V}(t)-p_{S+1}^{V}(t))+\varepsilon(1-p_{S+1}^{\ncc}(t)-2p_{S+1}^{V}(t))\nonumber\\
&&+\,\beta(1-p_{S+1}^{\ncc}(t)-p_{S+1}^{V}(t)).\nonumber
\end{eqnarray}
\end{lemm}

Note that equations~(7) and (8) do not depend on $\lambda_{1}$.

\remno{2} It can easily be checked that the  system~(\ref{eq:evol-dyrok}) 
has the same form as the corresponding system for the simple symmetric 
exclusion process on the finite set $0,1,\ldots,S+1$ (with empty 
boundary conditions).

The remaining part of this subsection is devoted to the proof of 
Lemma~\ref{l-zamk-odnoch}. First we shall obtain equations for  the 
marginal distributions $q_{t}[(k,x);e]$ {\it without assumption} 
of translation invariance of the initial distribution of velocities.

\label{sub:urav-marg}

For inner layers $1\le k\le S$ the marginal distributions satisfy  the
following equations.
\begin{eqnarray}
&&\!\!\!\!\frac{d}{dt}\,q_{t}[(k,x);\ncc]  = \lambda\big(q_{t}[(k+1,x);\ncc]+q_{t}[(k-1,x);\ncc]
-2q_{t}[(k,x);\ncc]\big)\nonumber \\
&&\quad +\,\lambda_{1}\big(q_{t}[((k,x),(k,x+1));(V,\ncc)]-q_{t}[((k,x-1),(k,x));(V,\ncc)]\big),
\label{eq:l1-k-ncc}\\
&&\!\!\!\!\frac{d}{dt}\,q_{t}[(k,x);0]  = \lambda\big(q_{t}[(k+1,x);0]+
q_{t}[(k-1,x);0]-2q_{t}[(k,x);0]\big)\nonumber \\
&&\quad +\,\varepsilon\big(q_{t}[(k,x);V]-q_{t}[(k,x);0]\big),\nonumber\\
&&\!\!\!\!\frac{d}{dt}\,q_{t}[(k,x);V]  = \lambda\big(q_{t}[(k+1,x);V]+q_{t}[(k-1,x);V]-2q_{t}[(k,x);V]
\big)\nonumber \\
&&\quad +\,\varepsilon\big(q_{t}[(k,x);0]-q_{t}[(k,x);V]\big)+\,\lambda_{1}\big(-q_{t}[((k,x),(k,x+1));(V,\ncc)]\nonumber\\
&&\quad\qquad+\,q_{t}[((k,x-1),(k,x));(V,\ncc)]\big).
\label{eq:l1-k-V}
\end{eqnarray}

On the bottom layer $k=0$ the equations have the following form
\begin{eqnarray}
&&\!\!\!\!\frac{d}{dt}\,q_{t}[(0,x);\ncc]  = \lambda\big(q_{t}[(1,x);\ncc]-q_{t}[(0,x);\ncc]\big)\nonumber\\
&&\quad +\,\lambda_{1}\big(q_{t}[((0,x),(0,x+1));(V,\ncc)]-q_{t}[((0,x-1),(0,x));(V,\ncc)]\big),
\label{eq:l1-0-ncc}\\
&&\!\!\!\!\frac{d}{dt}\,q_{t}[(0,x);0]  =  \lambda\big(q_{t}[(1,x);0]-q_{t}[(0,x);0]\big)+\beta q_{t}[(0,x);V]\nonumber \\
&&\quad +\,\varepsilon\big(q_{t}[(0,x);V]-q_{t}[(0,x);0]\big),\nonumber\\ &&\!\!\!\!\frac{d}{dt}\,q_{t}[(0,x);V]  = \lambda\big(q_{t}[(1,x);V]-q_{t}[(0,x);V]\big)-\beta q_{t}[(0,x);V]\nonumber \\
&&\quad +\,\varepsilon\big(-q_{t}[(0,x);V]+q_{t}[(0,x);0]\big)+\,\lambda_{1}\big(-q_{t}[((0,x),(0,x+1));(V,\ncc)]\nonumber\\
&&\qquad\quad+\,q_{t}[((0,x-1),(0,x));(V,\ncc)]\big).
\label{eq:l1-0-V}
\end{eqnarray}

On the top layer $k=S+1$ we have the equations
\begin{eqnarray}
\frac{d}{dt}\,q_{t}[(S+1,x);\ncc] & = & \lambda\big(q_{t}[(S,x);\ncc]-q_{t}[(S+1,x);\ncc]\big)
\label{eq:l1-S-1-ncc}\nonumber\\
&&+\,\lambda_{1}\big(q_{t}[((S+1,x),(S+1,x+1));(V,\ncc)]\nonumber\\
&&\qquad\ \ -\,q_{t}[((S+1,x-1),(S+1,x));(V,\ncc)]\big),\\
\frac{d}{dt}\,q_{t}[(S+1,x);0] & = & \lambda\big(q_{t}[(S,x);0]-q_{t}[(S+1,x);0]\big)+\beta q_{t}[(S+1,x);V]\nonumber \\
&& +\,\varepsilon\big(q_{t}[(S+1,x);V]-q_{t}[(S+1,x);0]\big),\nonumber\\
\frac{d}{dt}\,q_{t}[(S+1,x);V] & = & \lambda\big(q_{t}[(S,x);V]-q_{t}[(S+1,x);V]\big)+\beta q_{t}[(S+1,x);0]\label{eq:l1-S-1-V}\nonumber\\
&& +\,\varepsilon\big(q_{t}[(S+1,x);0]-q_{t}[(S+1,x);V]\big)\nonumber\\
&& +\,\lambda_{1}\big(-q_{t}[((S+1,x),(S+1,x+1));(V,\ncc)]\nonumber\\
&&\qquad\ \ +\,q_{t}[((S+1,x-1),(S+1,x));(V,\ncc)]\big).
\end{eqnarray}

We see that the equations for one-dimensional marginal distributions {\it 
are not closed}, since they contain two-dimensional  distributions in 
the r.h.s.

Derivation of these equations is  straightforward, but intermediate 
calculations are rather cumbersome. This derivation will be more 
transparent if we note that dynamics of the model consists of three 
processes. The first one is a symmetric exclusion process on vertical 
layers. The second component is a totally asymmetric exclusion process 
in the horizontal direction. And, finally, the Glauber spin-flip process at 
each point. This latter process enters the equations in a very simple 
manner, but  it is very useful to 
consider  the first two processes separately. We shall see that the equations for the first 
process are  closed even with the boundary conditions, while 
the equations for the second process are not closed.

{\it Model $1$\dvt symmetric exclusion process.} 
First it is convenient to 
consider a simpler  {\it auxiliary} model $\xi(t)=(\xi_{s}(t))$, which 
describes a single vertical layer  $x=0$, consisting of points 
$s=0,1,2,\ldots,S,S+1$, under additional assumption $\lambda_{1}=0$, 
$\varepsilon=0$ and $\beta=0$. For notational convenience we assume that 
there are no \otk holes\zaks, i.e.,\ $\xi_{s}(t)\in\{ 0,1\} $, and put 
$V=1$. Then the process $\xi(t)$, $t\ge0$, is a well-known simple 
exclusion process.

For $1\le i<i+k-1\le S$ and $e_{j}=0,1$ denote
$$
p_{t}(i;e_{1}\ldots e_{k})=\bP\,\{\xi_{i}(t)=e_{1},\ldots,\xi_{i+k-1}(t)=e_{k}\}.
$$
One-particle functions  $p_{t}(s;1)$ satisfy closed equations. Indeed, using an explicit form of the transition functions of the Markov process $\xi$ 
on a small time interval $[t,t+dt)$ and 
the complete probability formula, for any $1\le i\le S$  
 we have, up to terms of order  $o\,(dt)$,
\begin{eqnarray}
p_{t+dt}(i;1) & = & p_{t}(i-1;111)+[p_{t}(i-1;011)+p_{t}(i-1;110)](1-\lambda\, dt)\nonumber\\
&&+\,p_{t}(i-1;010)(1-2\lambda\,dt)+[p_{t}(i-1;100)\nonumber \\
&&+\,p_{t}(i-1;001)]\lambda\,dt+p_{t}(i-1;101)\cdot 2\lambda\, dt,\label{eq:korr-sym}
\end{eqnarray}
hence
\begin{eqnarray}
p_{t+dt}(i;1)-p_{t}(i;1) & = & \lambda\, dt\,[-p_{t}(i-1;01)-p_{t}(i;10)+p_{t}(i-1;10)+p_{t}(i;01)]
\nonumber \\
&=& \lambda\,dt\,[p_{t}(i-1;1)-2p_{t}(i;1)+p_{t}(i+1;1)],
\label{means}
\end{eqnarray}
and, finally,
\begin{equation}
\frac{d}{dt}\,p_{t}(i;1)=\lambda[p_{t}(i-1;1)-2p_{t}(i;1)+p_{t}(i+1;1)].
\label{eq:ur-m-sred}\end{equation}

{\it Model $2$\dvt asymmetric exclusion process.} 
It is convenient also to consider an asym\-met\-ric exclusion process 
$\zet(t)=(\zet_{x}(t),\, x\in\bZ)$ with three  values at the node: 
$\zet_{x}\in\{ \ncc,0,V\} $. We assume that  the following 
transitions are only possible: each pair of closest neighbors of the type 
$V\ncc$ become  $\ncc V$ with intensity $\lambda_{1}$. In other 
words, on the time interval $[t,t+dt]$ each particle, having velocity $V$,
independently of other particles jumps one unit to the right 
($x\rightarrow x+1$) with probability $\lambda_{1}dt+o\,(dt)$ provided 
that the node  $x+1$ is empty. We need to do calculations similar 
to~(\ref{eq:korr-sym}) and~(\ref{means}). Now  $x\in\bZ$.
As before, let us consider all possible states in the  neighboring nodes  
$x-1$,
$x$ and $x+1$ at time $t$ and apply the complete probability formula:
\begin{eqnarray*}
p_{t+dt}(x;\ncc) = \sum_{e_{i}\in\{ \ncc,0,V\} ,i=1,2,3}&&\!\!\!\bP\,\big(
\zet_{x}(t+dt)=\ncc\,|\,(\zet_{x-1}(t),\zet_{x}(t),\zet_{x+1}(t))=(e_{1},e_{2},e_{3})\big)\\
&&\!\!\!\times\,\bP\,\big\{(\zet_{x-1}(t),\zet_{x}(t),\zet_{x+1}(t))=(e_{1},e_{2},e_{3})\big\}.
\end{eqnarray*}
Conditional probabilities under the sum sign can be easily found. 
The results are presented in the 
following table, where   \otk $*$\zaks   means any of  symbols~$\ncc$,~$0$ or~$V$:

{\footnotesize
\renewcommand{\arraystretch}{0}
\newcommand{\strutdo}{\vrule depth0pt width0pt height10pt}
\newcommand{\strutposle}{\vrule depth5pt width0pt height0pt}
\newcommand{\strutdoposle}{\vrule depth5pt width0pt height10pt}
\begin{center}
\begin{tabular}{|ccc|c|}
\hline
\footnotesize\strutdo $x-1$&\footnotesize$x$&\footnotesize$x+1$&\footnotesize$\bP\,(\zet_{x}(t+dt)=\ncc\,|\,(\zet_{x-1}(t),\zet_{x}(t),\zet_{x+1}(t))=(e_{1},e_{2},e_{3}))$\\
\strut\footnotesize $e_{1}$&\footnotesize$e_{2}$&\footnotesize$e_{3}$&\\
\hline 
\rule{0pt}{2pt}&\rule{0pt}{0pt}&\rule{0pt}{0pt}&\rule{0pt}{2pt}\\
\hline
\strutdo \footnotesize$\ncc$&\footnotesize$\ncc$&\footnotesize$*$&\footnotesize$1-o\,(dt)$\\
\strut \footnotesize$0$&\footnotesize$\ncc$&\footnotesize$*$&\footnotesize$1-o\,(dt)$\\
\strutposle \footnotesize$V$&\footnotesize$\ncc$&\footnotesize$*$&\footnotesize$1-\lambda_{1}\,dt+o\,(dt)$\\
\hline 
\strutdoposle \footnotesize$*$&\footnotesize$0\footnotesize$&\footnotesize$*$&\footnotesize$o\,(dt)$\\
\hline 
\strutdo \footnotesize$*$&\footnotesize$V$&\footnotesize$\ncc$&\footnotesize$\lambda_{1}\,dt+o\,(dt)$\\
\strut \footnotesize$*$&\footnotesize$V$&\footnotesize$0$&\footnotesize$o\,(dt)$\\
\strutposle \footnotesize$*$&\footnotesize$V$&\footnotesize$V$&\footnotesize$o\,(dt)$\\
\hline
\end{tabular}
\par\end{center}
}

Up to terms of order $o\,(dt)$ we get
\begin{eqnarray*}
p_{t+dt}(x;\ncc) & = & p_{t}(x-1;\ncc\ncc)+p_{t}(x-1;0\ncc)+p_{t}(x-1;V\ncc)\\
&&-\,\lambda_{1}\,dt\cdot p_{t}(x-1;V\ncc)+\lambda_{1}\,dt\cdot p_{t}(x;V\ncc)\\
&=& p_{t}(x;\ncc)+\lambda_{1}\,dt\cdot[-p_{t}(x-1;V\ncc)+p_{t}(x;V\ncc)],
\end{eqnarray*}
and, hence,
\begin{equation}
\frac{d}{dt}\,p_{t}(x;\ncc)=\lambda_{1}[-p_{t}(x-1;V\ncc)+p_{t}(x;V\ncc)].\label{eq:ur-sr-assym}
\end{equation}

In the same way, we obtain the equation for  $p_{t}(x;V)$:
$$
\frac{d}{dt}\,p_{t}(x;V)=\lambda_{1}[p_{t}(x-1;V\ncc)-p_{t}(x;V\ncc)].
$$
Similarly one can consider the components of the dynamics~(\ref{eq:gen-L-obsch}),
corresponding to the random perturbations and to the behavior on the boundaries.

It appears that the equations for  $q_{t}[(k,x);e]$ become closed if we 
consider only translation invariant distributions. Namely, to finish the proof 
of Lemma~\ref{l-zamk-odnoch}, from now on we assume that the 
process starts at time  $t=0$ from a {\it translation invariant} 
distribution. Recall that the dynamics conserves the translation 
invariance, i.e. at any time  $t>0$ the process has a 
translation invariant distribution. In particular, 
the property~(\ref{eq:m1-m2-tr-inv}) holds. Therefore, in the right hand sides of the 
equations~(\ref{eq:l1-k-ncc})--(\ref{eq:l1-S-1-V}) all summands having 
the factor $\lambda_{1}$ disappear. Collecting together the equations with 
$\frac{d}{dt}\,q_{t}((k,x);\ncc)$, $0\le k\le S+1$, and using 
notation~(\ref{eq:p-e-s-t}) , we come to the  statement~1) of 
Lemma~\ref{l-zamk-odnoch}. The statement~2) easily follows from 
the form of the equations~(\ref{eq:l1-k-ncc})--(\ref{eq:l1-S-1-V}) if we 
take into account~(\ref{eq:normQ0V}).

This completes the proof of the lemma.

\subsection{Convergence as $t\rightarrow\infty$}
We prove here case~1) of Theorem~\ref{t-first}, namely,
we  show the existence of the limits $\lim_{t\rightarrow\infty}p_{k}^{e}(t)$
for the functions $p_{k}^{e}(t)$, satisfying  the 
equations~(\ref{eq:evol-dyrok})--(\ref{eq:evol-V}).
We  use probabilistic arguments.
 On the  {\it finite} state space 
$\ccM=\{\ncc,0,V\}^{\{ 0,1,\ldots,S+1\}}$ let us define
an auxiliary continuous time Markov process
$$
\auxpr(t)=(\auxpr_{k}(t),\ k=0,1,\ldots,S+1),\qquad\auxpr_{k}(t)\in\{\ncc,0,V\},
$$
with the following transitions. 
On a small time interval $[t,t+dt]$ the following events can occur 
independently of each other:

-- any nodes  $k$ and $k+1$  exchange their states with probability
$\lambda\, dt+o\,(dt)$; 

-- any node of the zero layer $k=0$, being in the state $V$, changes its 
state to  $0$ with probability $\beta\,dt+o\,(dt)$; any node of the top 
layer $k=S+1$, being in the state~$0$, with probability 
$\beta\,dt+o\,(dt)$ changes its state to~$V$; 

--~any node $k$ in a state different from $\ncc$ changes its state according to the rule
 $0\leftrightarrow V$ with probability $\varepsilon\, dt+o\,(dt)$ 
 ({\it random perturbation}). 

Note that the Markov process $\auxpr(t)$  is reducible, because its dynamics 
does not change the total number of nodes
having the empty state ($\ncc$). 
At the same time the process $\auxpr(t)$ considered on any set, invariant 
with respect to the dynamics 
$$
\ccM_{n}=\{\auxpr=(\auxpr_{0},\ldots,\auxpr_{S+1})\in\ccM\cl\#\big\{ j\cl\auxpr_{j}=\ncc\} =n\} ,\qquad n=0,1,\ldots,S+1,
$$
is irreducible and ergodic. Hence, for any  $n$  the following limits
$$
\lim_{t\rightarrow\infty}\bP\,(\auxpr(t)=\varkappa\,|\,\auxpr(0)=\varkappa_{0})
=g_{n}(\varkappa),\qquad\varkappa_{0},\varkappa\in\ccM_{n},
$$
exist and do not depend on the concrete choice of initial state~$\varkappa_{0}$.
Thus for any initial distribution 
  $\mu_{\auxpr(0)}$ there exists the  limit
 $\lim_{t\rightarrow\infty}\bP\,\{ \auxpr(t)=\varkappa\} $,
depending on $\mu_{\auxpr(0)}$.

If we permit, within the current subsection, the following notation
$$
p_{k}^{e}(t)=\bP\,\{ \auxpr_{k}(t)=e\} ,\qquad e\in\{ \ncc,0,V\} ,\quad k=0,1,\ldots,S+1,
$$
then it is easy to see that the functions $p_{k}^{e}(t)$ satisfy 
the system of equations~(\ref{eq:evol-dyrok})--(\ref{eq:evol-V}). 
Proof of this fact is even simpler than the reasonings of subsection~\ref{sub:urav-marg}.
Since
$$
p_{k}^{e}(t)=\sum_{\varkappa\in\ccM:\,\varkappa_{k}=e}\,\bP\,\{ \auxpr(t)=\varkappa\},
$$
the  statement~1) of Theorem~\ref{t-first} easily follows.

\subsection{Exact formulas for stationary solutions}
Now we return to the process  $v(t)$. Denote by
  $\mu_{s}^{\ncc}$,
$\mu_{s}^{0}$ and
 $\mu_{s}^{V}$ the stationary probabilities of the following three events:
 on the vertical layer
 $s$ there is no particle, 
 there is a particle with velocity
$0$ and 
there is a particle with velocity
 $V$ correspondingly, that is
  $\lim\limits_{t\rightarrow\infty}p_{s}^{e}(t)=\mu_{s}^{e}$.
As it was explained in the previous subsection these limits depend on the initial 
distribution~$\eta^{0}$. In particular, 
\begin{equation}
\sum_{s=0}^{S+1}\mu_{s}^{\ncc}=(S+2)-M(\eta^{0}),\label{eq:invar-1}
\end{equation}
where $M(\eta^{0})$ is defined in~(\ref{eq:M-eta-0}).

Let us remark that the limiting probabilities
 $\mu_{s}^{e}$ satisfy the stationary versions of the 
 equations~(\ref{eq:evol-dyrok}),~(\ref{eq:evol-V}).
It follows from~(\ref{eq:normQ0V}) that
$$
\mu_{s}^{\ncc}+\mu_{s}^{0}+\mu_{s}^{V}=1\qquad\forall s.
$$

\begin{lemm}
The distribution of \otk  holes\zaks is uniform on the set of layers\dvt 
$\mu_{s}^{\ncc}=\mu_{s_{1}}^{\ncc}$
for all $s,s_{1}\in\{ 0,1,\ldots,S+1\}$.
\end{lemm}

\Proof The system of stationary equations for $\mu_{s}^{\ncc}$,
$s=0,1,\ldots,S+1$, has the form:
\begin{eqnarray}
0 & = & \mu_{s-1}^{\ncc}+\mu_{s+1}^{\ncc}-2\mu_{s}^{\ncc},\qquad 1\le s\le S,\nonumber\\
0 & = & \mu_{1}^{\ncc}-\mu_{0}^{\ncc},\nonumber\\
0 & = & \mu_{S}^{\ncc}-\mu_{S+1}^{\ncc}.\nonumber
\end{eqnarray}
It is evident that the set of all solutions of this system is the one-dimensional subspace
$\mu_{0}^{\ncc}=\cdots=\mu_{S+1}^{\ncc}$. 
From the probabilistic nature of our model we have
$$
\mu_{0}^{\ncc}=\cdots=\mu_{S+1}^{\ncc}=1-\rho_{S}\in[0,1].
$$

This proves the lemma and the  statement~2) of Theorem~\ref{t-first}.

The number $\rho_{S}\in[0,1]$ will be called 
 particle density. 
It is readily seen from~(\ref{eq:invar-1}) that the density depends on the initial
distribution: 
$\rho_{S}=M(\eta^{0})/(S+2)$.

Since $\mu_{k}^{0}+\mu_{k}^{V}\equiv\rho_{S}$, 
we need only to find probabilities $(\mu_{k}^{V},\ k=0,1,\ldots,S+1)$.
On the inner layers they satisfy  the following equations\dvt
\begin{eqnarray}
0 & = & \lambda(\mu_{k+1}^{V}+\mu_{k-1}^{V}-2\mu_{k}^{V})+\varepsilon(\rho_{S}-2\mu_{k}^{V}),\qquad 1\le k\le S,\label{eq:razn-neodn}
\end{eqnarray}
with the following boundary conditions
\begin{eqnarray}
0 & = & \lambda(\mu_{1}^{V}-\mu_{0}^{V})+\varepsilon(\rho_{S}-2\mu_{0}^{V})
-\beta\mu_{0}^{V},\label{eq:kr-u-0}\\
0 & = & \lambda(\mu_{S}^{V}-\mu_{S+1}^{V})+\varepsilon(\rho_{S}-2\mu_{S+1}^{V})
+\beta(\rho_{S}-\mu_{S+1}^{V}).\label{eq:kr-u-S+1}
\end{eqnarray}

Case\ 1: $\varepsilon=0$.
Equations  (\ref{eq:razn-neodn}) take the following form
 $\mu_{k+1}^{V}+\mu_{k-1}^{V}-2\mu_{k}^{V}=0$. Characteristic equation corresponding to this
 difference equation 
$$
z^{2}+1-2z=0
$$
has a root $z=1$ of order~2. This is the {\it resonant} case, therefore a general solution 
is $\mu_{k}^{V}=D_{0}+D_{1}k$.
The coefficients $D_{0}$ and $D_{1}$ can be found from the  boundary 
conditions~(\ref{eq:kr-u-0}) and~(\ref{eq:kr-u-S+1}):
$$
0  =  \lambda D_{1}-\beta D_{0},\quad
0 = -\lambda D_{1}+\beta(\rho_{S}-D_{0}-D_{1}(S+1)).
$$
Solving the latter system, we get  explicit solution 
$$
\mu_{k}^{V}=\frac{\rho_{S}}{S+1+2\lambda\beta^{-1}}\,(k+\lambda\beta^{-1}),\qquad k=0,1,\ldots,S+1\, ,
$$
which is a linear function.
Now the statement of Theorem~\ref{t-eps-0} easily follows.

Case\ 2: $\varepsilon>0$.
It can be shown in a standard way that  general solution 
of the inhomogeneous difference equation~(\ref{eq:razn-neodn}) has the form
\begin{equation}
\mu_{k}=\frac{\rho_{S}}{2}+C_{1}z_{1}^{k}+C_{2}z_{2}^{k},\label{eq:muk-c1c2}
\end{equation}
where $z_{1}$ and $z_{2}$ are the roots of the characteristic equation
\begin{equation}
\lambda z^{2}+\lambda-2(\lambda+\varepsilon)z=0,\label{eq:har-ur}
\end{equation}
$C_{1}$ and $C_{2}$ are  unknown coefficients. Immediately, by solving 
equation~(\ref{eq:har-ur}), we get
\begin{equation}
z_{1,2}=1+\frac{\varepsilon}{\lambda}\pm\sqrt{\Big(1+\frac{\varepsilon}{\lambda}\Big)^{2}-1}.
\label{eq:z1-z2}
\end{equation}
Notice that if  $\varepsilon>0$ then the equation~(\ref{eq:har-ur}) has two
 {\it different} real roots: $z_{1}>1$ and $z_{2}<1$,
moreover, $z_{1}z_{2}=1$. Substituting~(\ref{eq:muk-c1c2}) to the boundary 
conditions~(\ref{eq:kr-u-0})--(\ref{eq:kr-u-S+1}), after some algebra we obtain the next
lemma.

\begin{lemm}\label{l-yavn-mu-V-k} The solution $(\mu_{s}^{V},\ s=0,1,\ldots,S+1)$
has the following explicit form\dvt
$$
\mu_{k}^{V}=\frac{\rho_{S}}{2}-\rho_{S}\,\frac{(\er(z_{1})z_{1}^{S+1-k}-\er(z_{2})z_{2}^{S+1-k})
-(\er(z_{1})z_{1}^{k}-\er(z_{2})z_{2}^{k})}{2(b^2(z_{1})z_{1}^{S+1}-b^2(z_{2})z_{2}^{S+1})},
$$
$k=0,1,\ldots,S+1,$ where $b(z):=1+({2\varepsilon+\lambda(1-z^{-1})})/{\beta}.$
\end{lemm}

\subsection{Asymptotics and phase transition}
Here we  consider  $\lambda$ and $\beta$ as fixed  parameters and assume that
 $\varepsilon\rightarrow0$, $S\rightarrow\infty$. 
We see that for small~$\varepsilon$
$$
z_{1} = 1+\sqrt{\frac{2\varepsilon}{\lambda}}+O\Big(\frac{\varepsilon}{\lambda}\Big),\quad
z_{2} = 1-\sqrt{\frac{2\varepsilon}{\lambda}}+O\Big(\frac{\varepsilon}{\lambda}\Big).
$$

Taking into account the assumptions of Theorem~\ref{t-turb}, for simplicity we put 
 $\rho_{S}\equiv\rho$.

We are interested in the behaviour of
 $(\mu_{k}^{V},\,0\le k\le S+1)$
on the space scale $k=[uS],$ $u\in[0,1]$. To get a meaningful asymptotics 
in the limit $S\rightarrow\infty$, we assume that the parameter
$\varepsilon$ is a function of $S$:
$$
\varepsilon_{S}=\frac{\lambda}{2}\psi^{2}(S)S^{-2},
$$
where $\psi(S)>0$ is such that $\psi(S)S^{-1}\rightarrow 0$ as
 $S\rightarrow\infty$. Hence, $\sqrt{{2\varepsilon_{S}}/{\lambda}}=\psi(S)S^{-1}$ 
 and 
\begin{eqnarray*}
z_{1,2} & = & 1\pm\psi(S)S^{-1}+O(\psi^{2}(S)S^{-2}),\\
b(z_{1,2}) & = & 1\pm\frac{\sqrt{\lambda}}{\beta}\psi(S)S^{-1}+\beta^{-1}O(\psi^{2}(S)S^{-2}).
\end{eqnarray*}
The calculations presented below  show that the asymptotics of
 $\mu_{[uS]}^{V}$ strongly depends on the properties of the function~$\psi$.

In  case of {\it small} random perturbations $\psi(S)\rightarrow0$ 
\begin{eqnarray*}
\mu_{[uS]}^{V}&\sim&\frac{\rho}{2}-\rho\,\frac{2(1-u)\psi(S)-2u\psi(S)}{2\cdot2\psi(S)}+\beta^{-1}O(S^{-1})\\
&\sim&\frac{\rho}{2}+\rho\frac{2u-1}{2}=u\rho,\qquad u\in(0,1).
\end{eqnarray*}
and the profile has a laminar character.

In  case of {\it strong}  random perturbations $\psi(S)\rightarrow\infty$
\begin{equation}
\mu_{[uS]}^{V}\sim\frac{\rho}{2}-\rho\,\frac{z_{2}^{(1-u)S}-z_{2}^{uS}}{2z_{2}^{S}}+
\beta^{-1}O(S^{-1}\cdot\psi(S)e^{-\psi(S)})\rightarrow\frac{\rho}{2}
\qquad\forall u\in(0,1).\label{eq:siln-vozm}
\end{equation}
and the limiting profile is very far from being laminar. Indeed, it can be 
checked that the limiting profile here is the same
as in the case when there is no boundary conditions (formally, when $\beta=0$), that is called  free boundary conditions. 
We can conclude that if the  random perturbations are strong then the boundaries do not influence the inner layers. This is typical for the turbulent phase.

The case  $\psi(S)=\const$ {\it separates} the above two  {\it phases}.
For convenience let us put
 $\psi(S)=K,$ $K>0$. Then
\begin{eqnarray*}
\mu_{[uS]}^{V}&\sim&\frac{\rho}{2}-\rho\,\frac{(e^{K(1-u)}-e^{-K(1-u)})-(e^{Ku}-e^{-Ku})}{2(e^{K}-
e^{-K})}\\
&=&\frac{\rho}{2}\Big(1+\frac{\sh(K(u-1/2))}{\sh(K/2)}\Big),\qquad u\in(0,1).
\end{eqnarray*}

Let us remark in conclusion that the assumption  $\psi(S)S^{-1}\rightarrow0$ 
is not essential for the strong perturbation asymptotics ~(\ref{eq:siln-vozm}).
In particular, this results holds in the important case
 $\varepsilon_{S}=\varepsilon$.
Indeed, in this case $z_{2}$ does not depend on $S$, and, using the
explicit formula of Lemma~\ref{l-yavn-mu-V-k}, we get for any $u\in(0,1)$
$$
\mu_{[uS]}^{V}-\frac{\rho}{2}\sim\rho\,\frac{\const}{b(z_{2})z_{2}^{(1-\min(1-u,u))S}}\rightarrow 0\qquad(S\rightarrow\infty).
$$

We stress that all limiting expressions do not depend on
the parameter~$\beta$.

Theorem~\ref{t-turb} is proved.


\begin{thebibliography}{99}

\bibitem{Strange}{\em Strange attractors}. Mir Moscow, 1981. (Mathematics. News of foreign science, V.~22.)

\bibitem{ChorinMarsd} {\sc Chorin~A.\,J.\kb Marsden~J.\,E.} {\em A Mathematical Introduction to Fluid Mechanics}. New York: Springer-Verlag, 1993, 169~p.

\bibitem{MarPul} {\sc Marchioro~C.\kb Pulvirenti~M.} {\em Mathematical Theory
of Incompressible Non Viscous Fluids}. New York: Springer-Verlag. 1994, 283~p.

\bibitem{Shir} {\sc Shiryaev~A.} {\em On the classical, statistical, and stochastic approaches to the hydrodynamic turbulence}. Research Report \No 2. Aarhus: Thiele Centre for Applied Mathematics in Natural Science, 2007.

\bibitem{Kolmogor} {\sc Kolmogorov A.N.} {\em Mathematical models of turbulent motion of an incompressible viscous fluid}.
 RUSS MATH SURV, 2004, v.~59,  \No 1, p.~3-9. 
 


\bibitem{MeshSin} {\sc Meshalkin L.\,D.\kb and Sinai Ia.\,G.}
 {\em Investigation of the stability of a stationary solution of a system of equations 
 for the plane movement of an incompressible viscous liquid}. 
 Journal of Applied Mathematics and Mechanics, 1961,
 v.~25,   \No  6,  p.~1700-1705 

\bibitem{LanAxiLif} {\sc Landau L.\,D.\kb Akhiezer A.\,I.\kb Lifshitz E.\,M.} 
{\em General Physics:  Mechanic and Molecular Physics}. Pergamon Press, 1970, 372~p.

\bibitem{Liggett} {\sc Liggett~T.} {\em Interacting Particle Systems}.
Springer, 2005, 496~p.

\bibitem{Kipnis} {\sc Kipnis~C.\kb Landim~C.} {\em Scaling Limits of Interacting
Particle Systems}. Berlin: Springer-Verlag, 1999, 442~p.

\bibitem{Spohn} {\sc Spohn~H.} {\em Large Scale Dynamics of Interacting Particles}.
Berlin: Springer-Verlag. 1991, 342~p.

\bibitem{EsMaYao} {\sc Esposito~R.\kb Marra~R.\kb Yau~H.\,T.} 
{\em Navier--Stokes equations for stochastic particle
systems on the lattice}. Comm.\ Math.\ Phys., 1996, v.~182, \No 2, p.~395--456.

\bibitem{IMM} {\sc Ignatyuk~I.\,A.\kb Malyshev~V.\,A.\kb Molchanov~S.\,A.} {\em Moment-closed processes with local interaction}. Selecta Math.\ Soviet., 1989, v.~8, \No 4, p.~351--384.

\bibitem{Meleard} {\sc M\'el\'eard~S.} {\em Monte-Carlo approximations for 2d Navier--Stokes equations with measure initial data}. Probab.\ Theory Related Fields, 2001, v.~121, \No 3, p.~367--388.

\bibitem{Fontbona} {\sc Fontbona~J.} {\em A probabilistic interpretation and stochastic particle approximations of the 3-dimensional Navier--Stokes
equations}. Probab. Theory Related Fields, 2006, v.~136, \No 1, p.~102--156.

\bibitem{Golinelli} {\sc Golinelli~O.\kb Mallick~K.} {\em Derivation of a matrix product representation for the asymmetric exclusion process from algebraic
Bethe ansatz}. J.\ Phys.\ A., 2006, v.~39, \No 34, p.~10647--10658.

\bibitem{Derrida07} {\sc Derrida~B.\kb Lebowitz~J.\,L.\kb Speer~E.\,R.} {\em Entropy of open lattice systems}. J.~Statist. Phys., 2006, v.~126, \No 4/5,
p.~1083--1108. 

\end{thebibliography}
\end{document}